\documentclass[lettersize,journal]{IEEEtran}
\usepackage{amsmath,amsfonts}
\usepackage{algorithmic}
\usepackage{algorithm}
\usepackage{array}
\usepackage{textcomp}
\usepackage{url}
\usepackage{verbatim}
\usepackage{graphicx}
\usepackage{cite}
\usepackage{xcolor}

\usepackage{subcaption}
\usepackage{multicol}
\usepackage{tcolorbox}
\usepackage[colorlinks = true,
            linkcolor = black,
            urlcolor  = blue,
            citecolor = black,
            anchorcolor = blue]{hyperref}
            
\usepackage[noabbrev,capitalise]{cleveref}
\usepackage{tcolorbox}
\tcbuselibrary{listings,skins}
\usepackage{listings}
\usepackage{xcolor}
\definecolor{diffadd}{RGB}{220,255,220}
\definecolor{diffrem}{RGB}{255,220,220}

\newcommand{\red}[1]{\textcolor{red}{#1}}
\newcommand{\RCTR}{ARCTR}
\begin{document}

\title{\vspace{0in}Yet Another Mirage of Breaking MIRAGE:\\ \LARGE{Debunking Occupancy-based Side-Channel Attacks on Fully Associative Randomized Caches}}
\date{}

\author{
\begin{minipage}[t]{0.49\textwidth}
\centering
Chris Cao \\
University of Toronto \\
{\normalsize \texttt{chrisj.cao@mail.utoronto.ca}}
\end{minipage}
\hfill
\begin{minipage}[t]{0.49\textwidth}
\centering
Gururaj Saileshwar \\
University of Toronto \\
{\normalsize \texttt{gururaj@cs.toronto.edu}}
\end{minipage}
\vspace{-0.35in}
}


\maketitle


\begin{abstract}
Recent work presented at USENIX Security 2025 (SEC'25) claims that occupancy-based attacks can recover AES keys from the MIRAGE randomized cache.
In this paper, we examine these claims and find that they arise from a modeling flaw in the SEC'25 paper. 
Most critically, the SEC'25 paper's simulation of MIRAGE uses a constant seed to initialize the random number generator used for global evictions in MIRAGE, causing every AES encryption they trace to evict the same deterministic sequence of cache lines.
This artificially creates a highly repeatable timing pattern that is not representative of a realistic implementation of MIRAGE, where eviction sequences vary randomly between encryptions.
When we instead randomize the eviction seed for each run, reflecting realistic operation, the correlation between AES T-table accesses and attacker runtimes disappears, and the attack fails.
These findings show that the reported leakage is an artifact of incorrect modeling, and not an actual vulnerability in MIRAGE.

\end{abstract}
\section{Introduction}
MIRAGE~\cite{MIRAGE} is a randomized cache design, proposed in 2021, that emulates a fully associative cache with globally random evictions,  eliminating set-conflict cache side channels.
It builds on theoretical foundations such as multiple randomized set indexing functions using block ciphers, and power-of-two-choices~\cite{power-of-2-choices} based load-balancing, guaranteeing that set-associative evictions are practically impossible in a system’s lifetime.
Given these strong guarantees, several works have examined whether MIRAGE’s security holds in practice.

In 2023, ``\textit{Are Randomized Caches Truly Random''} (ARCTR)~\cite{randcachebroken} claimed to induce set-conflicts in MIRAGE, breaking its security guarantees. However, subsequent work~\cite{saileshwar2023mirage} showed that these were the result of incorrect modeling by ARCTR, caused by a buggy cipher implementation, and that MIRAGE's security guarantees remained intact. 

More recently, the SEC 2025 paper, ``\textit{Systematic Evaluation of \underline{R}andomized Cache Designs against \underline{C}ache \underline{O}ccupancy''} (RCO)~\cite{chakraborty2025occ}, claims that MIRAGE is vulnerable to cache-occupancy-based side-channel attacks that can leak secret AES keys.
Specifically, RCO (in Section 7), claims that the AES T-Table implementation can leak the AES key on MIRAGE via the cache occupancy side-channel, and that MIRAGE’s fully associative eviction policy makes it more susceptible than other randomized caches.
A subsequent paper at SEC 2025, ``\textit{SoK: So, You Think You Know All About Secure Randomized Caches?}''~\cite{bhatlasok} reiterates these claims in its Figure 17.
This paper examines whether these claims hold up or whether they are artifacts of modeling flaws like the ARCTR paper.

First, we attempted to reproduce the AES key recovery results reported in the RCO paper using their artifact\footnote{We use the the code artifact with RCO's paper~\cite{chakraborty2025occ}  - \url{https://zenodo.org/records/14869981} (Github release \href{https://github.com/NimishMishra/randomized_caches/releases/tag/v2.0}{v2.0}, commit: fc07ea6).}, specifically the guessing entropy (GE) for an unknown victim key (Figure 10 in the RCO paper~\cite{chakraborty2025occ}).
Using the RCO artifact, we initially could not reproduce their results.

\begin{figure}[h]
    \centering
    \vspace{-0.1in}
    \includegraphics[width=\columnwidth]{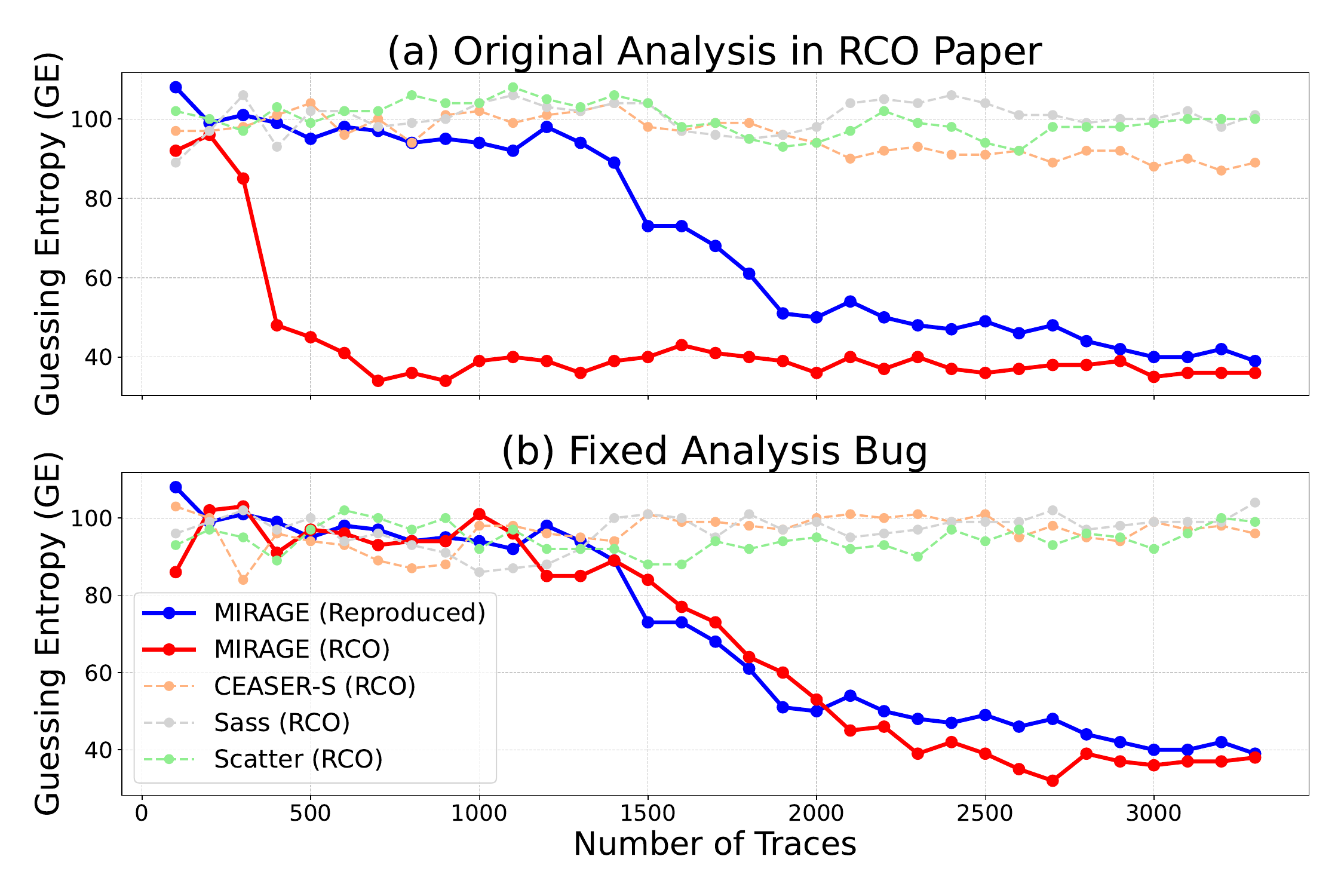}
    \vspace{-0.3in}
    \caption{\textbf{Guessing Entropy (GE) for an unknown AES key}, as the number of AES encryption traces increases. (a) Original analysis from RCO shows GE reduction for MIRAGE after 300 traces. (b) After fixing the analysis bug, GE reduction occurs after 1400 traces, consistent with our reproduction.}
   \vspace{-0.1in} 
    \label{fig:naive_reproduction}

\end{figure}

\textbf{Bug-1: GE Analysis Bug in RCO.} \cref{fig:naive_reproduction} compares the guessing entropy (GE) for an AES key (higher is better) for MIRAGE using RCO’s raw data (RCO – \red{red} line) and our reproduced data (Reproduced – \textcolor{blue}{blue} line), along with other randomized caches.
In the original analysis (\cref{fig:naive_reproduction}a), RCO reports GE dropping below 80 after just 300 AES traces. In contrast, our reproduced data shows the same GE reduction occurs after 1400 traces.

Upon investigation, we identified a \textit{bug} in the RCO artifact’s GE analysis code. The bug stems from an incorrect indexing of AES traces used for the analysis (see \cref{app:bug1}), which are spread over \textit{six} trace files in RCO's raw data, due to which their code uses roughly 6$\times$ more traces than intended to calculate GE. This results in an under-reporting of the number of traces required to reach a given GE in the RCO paper. After fixing this bug in RCO's artifact\footnote{The code for our reproduction, with our bug fixes, is open-sourced at: \url{https://github.com/sith-lab/yet-another-mirage-of-breaking-mirage}}, the analysis using RCO’s raw data matches our naive reproduction (\cref{fig:naive_reproduction}b), with GE for MIRAGE now dropping after 1400 traces.

Next, we examine the validity of this leakage.

\textbf{Bug-2: Global Eviction Modeling Bug in RCO.} 
RCO attributes AES key leakage to last-round S-Box accesses in the AES encryption influencing cache occupancy.
The attacker observes this by measuring access times to its own cached array, building templates for each key-byte value using a profiled key, and matching these templates against timings for an unknown key to recover it.
However, this explanation ignores MIRAGE’s \textbf{random global evictions}, which introduce noise in cache occupancy.
MIRAGE randomly evicts an existing LLC line upon every new insertion. Hence, repeated AES encryptions of even the same plaintext and key evict different addresses and produce different occupancy (unrelated to the key), and the attacker’s own accesses also evict its addresses randomly on each run, introducing further noise. Thus, an attacker would even find fingerprinting a key difficult (\cref{fig:debunk_exectime}), let alone leak the entire key byte-wise.
Investigating this, we uncover a second \textit{bug} in RCO’s evaluation:

\begin{tcolorbox}[boxrule=1pt,left=5pt,right=5pt,top=3pt,bottom=3pt]
\textbf{Modeling Flaw in RCO.} 
RCO’s simulations initialize the RNG used by MIRAGE’s evictions with a \textbf{static seed} before each AES encryption, causing a \textbf{fixed} sequence of evictions on each AES run, not practical in a real attack.
\end{tcolorbox}

RCO’s simulations seed MIRAGE’s RNG used for evictions with a \textbf{fixed value (42)} before each AES encryption, restarting each AES run from an identical state (see \cref{app:bug2}).
This produces a fixed eviction sequence on each AES encryption, not representative of MIRAGE.
In a realistic modeling of MIRAGE with random global evictions, the final cache occupancy (\textbf{O}) after each AES encryption should be determined by both the victim’s accesses (\textbf{V}) and the global random eviction sequence (\textbf{R}), $O\approx f(R,V)$. However, as RCO uses a constant RNG seed, producing a fixed eviction sequence in each run, the cache occupancy becomes a function of just victim accesses, $O \approx f(V)$, artificially creating correlations with the key.

To correctly model MIRAGE while restarting the simulation for each AES run, we \textbf{randomize the eviction RNG seed for each AES run} (e.g., using \texttt{std::random\_device}\footnote{Note that on some systems without a hardware-based entropy source, \texttt{std::random\_device} can also result in a deterministic output~\cite{random_device}. Please validate that the source of randomness you use is truly random on your system.} to seed the RNG).
This models a realistic setting, where the attacker cannot reset the RNG to a constant state on each AES run.
With this correction, correlations between attacker timings and keys disappear, 
and the victim AES key's guessing entropy remains high (above 90\%), as shown in  \textbf{\cref{fig:final_reproduction}}.




\begin{tcolorbox}[boxrule=1pt,left=5pt,right=5pt,top=3pt,bottom=3pt]

\textbf{Results with Correct Modeling.} After randomizing the RNG seed for MIRAGE in each AES run, the victim key's guessing entropy (GE) remains high ($>$90\%), see \textbf{\cref{fig:final_reproduction}}. Thus, AES key leakage on MIRAGE is infeasible. 
\end{tcolorbox}

\smallskip

\noindent \textbf{To summarize, we make the following contributions:}
\begin{enumerate}
    \item We show that the AES key guessing entropy remains high in MIRAGE, contradicting the RCO paper’s claims.
    \item We identify that RCO's usage of a fixed RNG seed for MIRAGE's evictions creates artificial correlations;  randomizing the seed eliminates any observed leakage. 
\end{enumerate}

\section{Background}
\subsection{The MIRAGE Cache}
MIRAGE~\cite{MIRAGE} is a randomized cache that prevents conflict-based side channels by emulating a fully-associative randomized cache: every eviction is global and selected uniformly at random from the entire cache. To support this, MIRAGE over-provisions invalid tags in each set and uses load-balancing via the Power-of-2-Choices to maintain available space, avoiding set-associative evictions. On a miss, the data-store victim is chosen randomly from all cache lines, its tag is located via a Reverse Pointer (RPTR) and removed, and the new tag is inserted via a Forward Pointer (FPTR). 
With an 8-way cache augmented with 6 extra ways (75\% extra) in the tag-store, MIRAGE guarantees the probability of a set-associative eviction is once in $10^{34}$ cache installs, an event that takes $10^{17}$ years to occur, making set-conflicts practically impossible in a system’s lifetime and eliminating conflict-based attacks.

\subsection{Cache Occupancy Attacks on Mirage}
Cache occupancy attacks measure changes in the overall occupancy of a shared cache, rather than targeting specific sets as in set-conflict attacks like Prime+Probe. Because they exploit aggregate cache usage, all randomized caches without explicit cache partitioning, such as CEASER, ScatterCache, and Mirage in principle, leak some information via cache occupancy. In fact, MIRAGE’s threat model explicitly claims to not protect against such occupancy-based channels. While covert channels between two colluding processes can be naively constructed by modulating the cache occupancy, the RCO~\cite{chakraborty2025occ} paper claims the existence of a stronger side-channel on MIRAGE: leaking AES keys via cache occupancy effects. 

The RCO paper claims that a victim using an AES T-Table implementation, can be forced to leak the key in MIRAGE, by a spy first priming the LLC to a chosen occupancy, letting the victim run one encryption, then timing accesses to the attacker’s own cache lines. By correlating these timings with last-round T-table accesses for guessed keys, they report low guessing entropy for AES keys (lower than 30\%), and claim full 128-bit AES key recovery on MIRAGE within a few hours.
This paper examines these claims of RCO.

\begin{figure*}[htbp]
    \centering
    \begin{subfigure}{0.49\textwidth}
        \centering
        \includegraphics[width=1\linewidth]{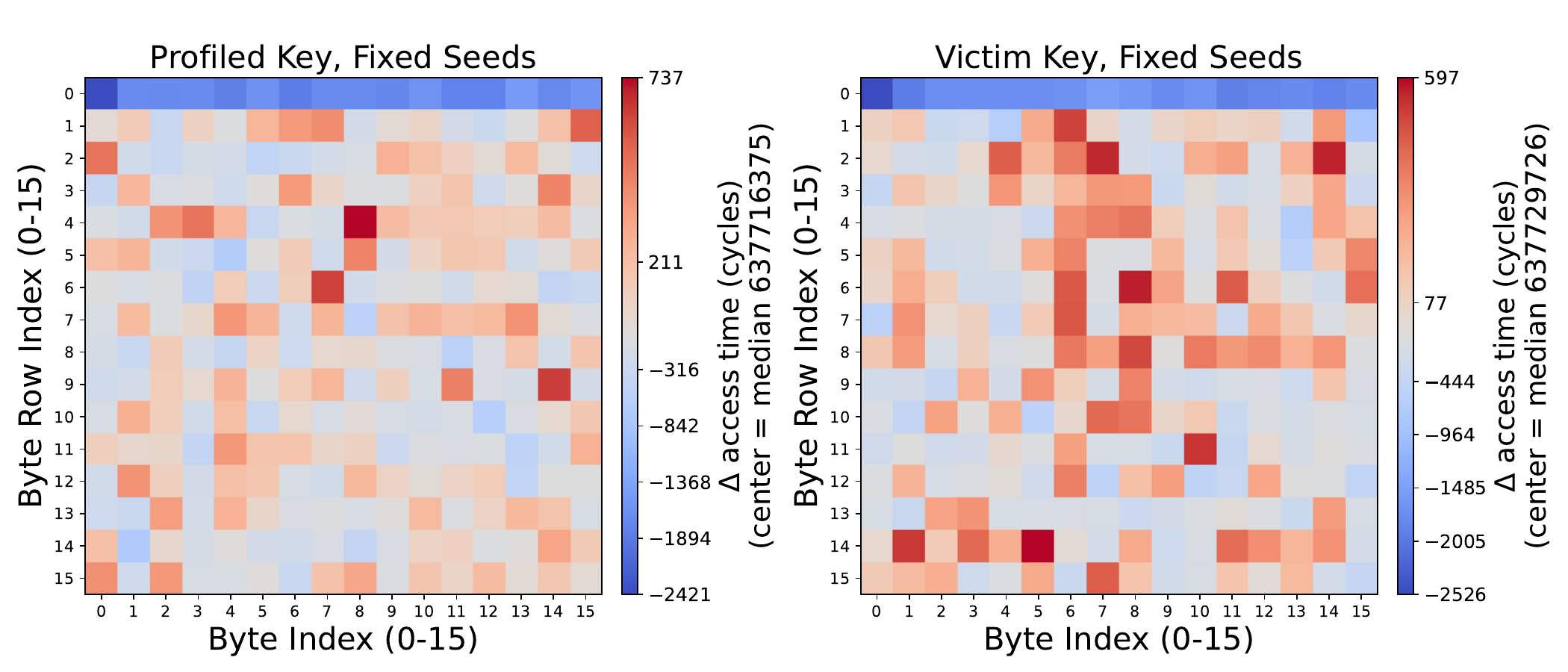}
        \caption{Fixed Seed for Global Evictions (RCO~\cite{chakraborty2025occ})}
        \label{fig:heatmap-fixed}
    \end{subfigure}
    \hfill
    \hspace{-0.2in}
    \begin{subfigure}{0.49\textwidth}
        \centering
        \includegraphics[width=1\linewidth]{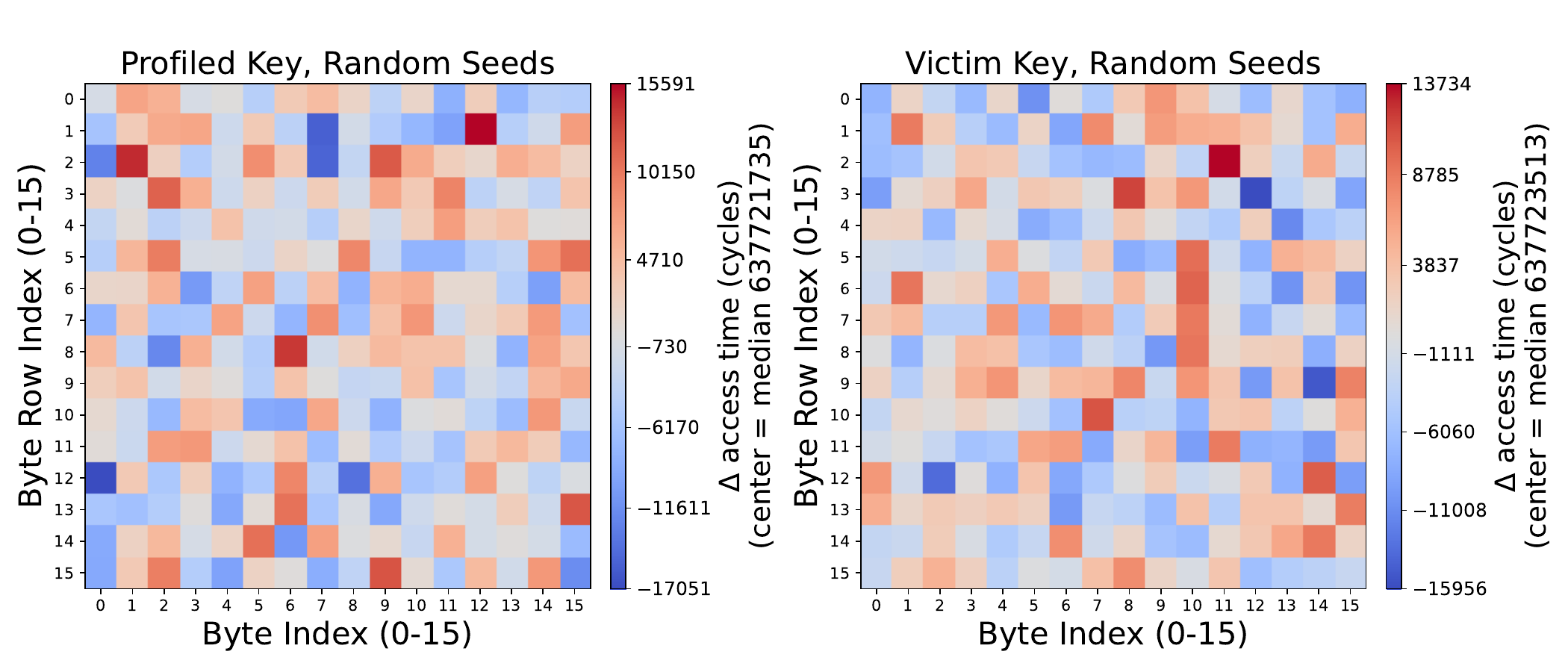}
        \caption{Random Seed for Global Evictions (our fix)}
        \label{fig:heatmap-rand}
    \end{subfigure}
    \caption{Heatmap of access times for the attacker to iterate through its array. We bin the access times, based on the 256 possible T-table entries accessed in the last round. The 256 bins are represented in the 16 x 16 matrix. (a) With Fixed Seed for Global Eviction, as used in RCO~\cite{chakraborty2025occ}, there is strong correlation between the heatmaps for profiled key and  victim key. (b) After our fix, with Random Seeds for Global Eviction, the correlations disappear, showing it is infeasible to  guess victim AES keys.}
    \label{fig:heatmaps}
\end{figure*}

\begin{figure*}[h]
    \centering
    \includegraphics[width=2\columnwidth]{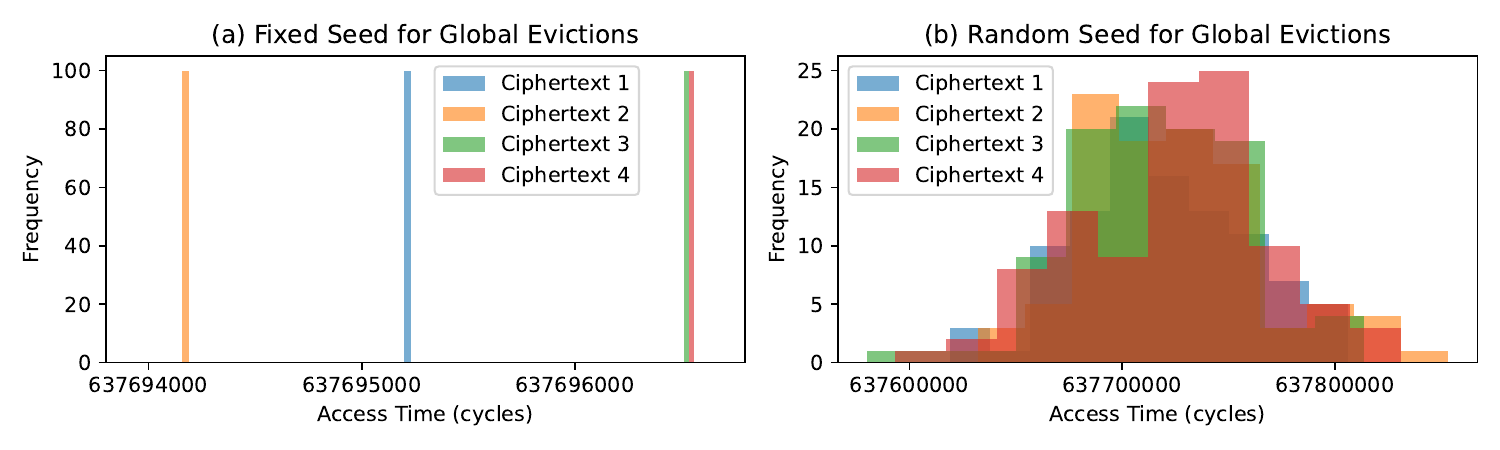}
     \caption{Access Times for the attacker with (a) original RCO implementation and (b) our bug fix for four sample plaintext-ciphertext pairs with the same key. (a) The original implementation uses a Fixed Seed (42) for Global Evictions in each AES encryption. (b) Our bug fix initializes the RNG used for Global Evictions, with a random seed (e.g., using \texttt{time()}), for each AES encryption, mimicking a real system where the seed cannot be reset to a static value each time. Once we address this issue, the different encryptions are indistinguishable based on access times to the attacker's array.}
     \vspace{-0.2in}
    \label{fig:debunk_exectime}
\end{figure*}
\section{Analyzing Claims of Occupancy-Based Side-Channel Attacks on MIRAGE}\label{sec:guarantees}


Using the authors' publicly released artifact, we first reproduced the results in Figure~10 of the RCO paper, as shown in \cref{fig:naive_reproduction}. 
Once we were able to reproduce the trend, that the guessing entropy for an unknown AES key reduces after 1400 encryptions, we analyze the root cause of the attack: that the execution time for an attacker accessing its own array is a function of the cache occupancy of the AES encryption, which is in turn a function of the victim's AES key.

\subsection{Pitfall-1: Fixed Sequence of Evictions Modeled by RCO}

\textbf{RCO Attack Root Cause.} The RCO~\cite{chakraborty2025occ} paper claims that the T-Table implementation of AES running on MIRAGE~\cite{MIRAGE} can leak the secret key through a cache occupancy attack.
The root cause of the leakage claimed by RCO is that the sequence of T-Table accesses in the last round of AES encryption, which depends on the secret key, can impact the cache occupancy in MIRAGE.
Therefore, the execution time for an attacker to access a large cached array, parts of which may have been evicted by the AES encryption, can leak the cache occupancy, and therefore the secret key.

\textbf{RCO Attack Mechanism.} To perform this attack,  RCO~\cite{chakraborty2025occ} creates a template for the execution times with all possible T-table entries accessed in the last round ($T_1$ to $T_{255}$), by using a known key (\textit{profiled key}).
The attacker measures the execution time to access its own cached array that occupies 50\% of the MIRAGE cache, using randomly generated plaintext-ciphertext pairs, and creates the template ($T_1$ to $T_{255}$) by averaging the access times for $T_n$= $\text{SBOX-Inv}(K \oplus CT )$, where \textit{K} and \textit{CT} are bytes of the last-round key and ciphertext. 
Such templates can be created for each of the bytes of the round-key (0 to 15). 
Later, for an unknown victim key, by creating a similar template using a guessed key and random plaintext-ciphertext pairs, the attacker identifies likely key values having the highest correlation with the profiled template.

\textbf{Reproducing RCO's Root-Cause.} 
To validate RCO's leakage, we try to reproduce its root cause, the templates with correlation between profiled and victim keys.
\cref{fig:heatmap-fixed} shows a heatmap visualizing the template built by the attacker, where each entry represents one of the 256 possible T-table entries (arranged as a 16$\times$16 matrix) and the cell color encodes the attacker’s average access time, when that entry is accessed in the last round of AES by the victim. If the profiled-key heatmap (attacker’s template) closely matches the heatmap with the guessed victim-key, the guess is likely to be the correct key.
For simplicity, instead of $T_n=\text{SBOX-INV}(K \oplus CT)$, we use $T_n= K \oplus CT$ for our bins, since SBOX-INV is just a lookup table, and visualize a single heatmap, averaging the heatmaps of key bytes 0 to 15. 

As shown in \cref{fig:heatmaps}(a), when we generate the heatmaps by using the RCO artifact~\cite{chakraborty2025occ} there is a clear correlation between the templates of the profiled key and the guessed victim key. This correlation can allow the attacker to leak the victim key since the templates will be highly correlated when the guessed key is actually the correct victim key.

\textbf{RCO's Bug: Fixed Sequence of Global Evictions.} 
To investigate the source of the observed correlation, we measured attacker access times after victim AES encryptions for four randomly chosen plaintext-ciphertext pairs, each repeated 100 times. Each encryption has a distinct last-round AES T-Table access sequence.
Figure~\cref{fig:debunk_exectime}(a) shows the histogram of attacker access times. CT1, CT2 and CT3 differ in access times (although CT3 and CT4 overlap), highlighting that the access times are indeed correlated with T-Table accesses. However, notably, all 100 repetitions of each encryption produce identical access-time measurements, despite global evictions being intended as random.

This reveals a bug in RCO: the global eviction RNG is seeded with a static value (42).
As a result, each AES execution follows the same fixed eviction sequence, making MIRAGE’s behavior deterministic. In a real implementation, the attacker cannot reset the RNG between runs, so the  deterministic behavior is unrealistic and artificially creates correlations between access times and AES T-Table sequences.

\smallskip
\textbf{Our Fix: Accurate Modeling with Random Seed.}
The cache occupancy ($O$) in MIRAGE is a function of both the Victim accesses ($V$) and the Global Eviction decisions ($GE$), i.e., $O \approx f(V, GE)$. RCO incorrectly assumes a fixed sequence of GE for each encryption, making the fingerprinting of $V$ using $O$ measurements possible.
We fix this bug in RCO, by changing the fixed seed for global evictions to a randomly chosen seed for each AES encryption simulation, representative of real systems, where the Global Eviction is performed by a RNG whose state cannot be reset by the attacker. After this bug fix, we see that the $O$ is now strongly impacted by the $GE$ in addition to the $V$.
As shown in \cref{fig:debunk_exectime}(b), with a random seed for each encryption, the attacker's access times show random variations of the order of 100,000 cycles, due to the global evictions unpredictably evicting the attacker’s own lines during its measurement phase. These overwhelm minor occupancy differences caused by the victim which varied timings by few 1000 cycles in \cref{fig:debunk_exectime}(a). 

After our fix, when we use a random seed for each AES encryption in \cref{fig:heatmaps}(b), the correlation between profiled and victim keys disappears, making the heatmaps virtually unrelated, eliminating the signal required for key leakage.
This realistic modeling of MIRAGE’s global evictions removes the attack’s timing signals, preventing AES key leakage on MIRAGE via occupancy attacks.
We provide the code for this bug fix in \cref{app:bug2}.




\subsection{Pitfall-2: Unrealistic L1 Cache Configuration}

The RCO paper claims to use a 512 \textbf{kilobyte} L1 cache, as per Section 7.1 of their paper~\cite{chakraborty2025occ}. However, their code artifact\footnote{We refer to the code artifact version 2.0 released by the authors on Zenodo  - \url{https://zenodo.org/records/14869981}} uses a 512 \textbf{byte} L1 cache by default which is not representative of real systems which typically possess atleast 64KB L1 Cache. Modeling a smaller L1 cache, such as 512 byte L1 cache, can inflate the L1 cache misses and LLC accesses, compared to a 64KB L1 Cache which can provide hits for \textit{all} T-table accesses after the first round. Thus, a 512 byte L1 cache can overestimate  the attack success.
We confirmed that after updating the initialized seed to be random, regardless of the L1 cache size being 512-byte like in the RCO artifact or a more realistic 64KB, there is a lack of correlation between the heatmaps of the profiled and victim key templates, similar to \cref{fig:heatmaps} (b), indicating that AES key leakage is impractical in MIRAGE.

\subsection{Guessing AES Key after Fixing RCO's Modeling Issues}

\begin{figure}[h]
    \centering
    \includegraphics[width=\columnwidth]{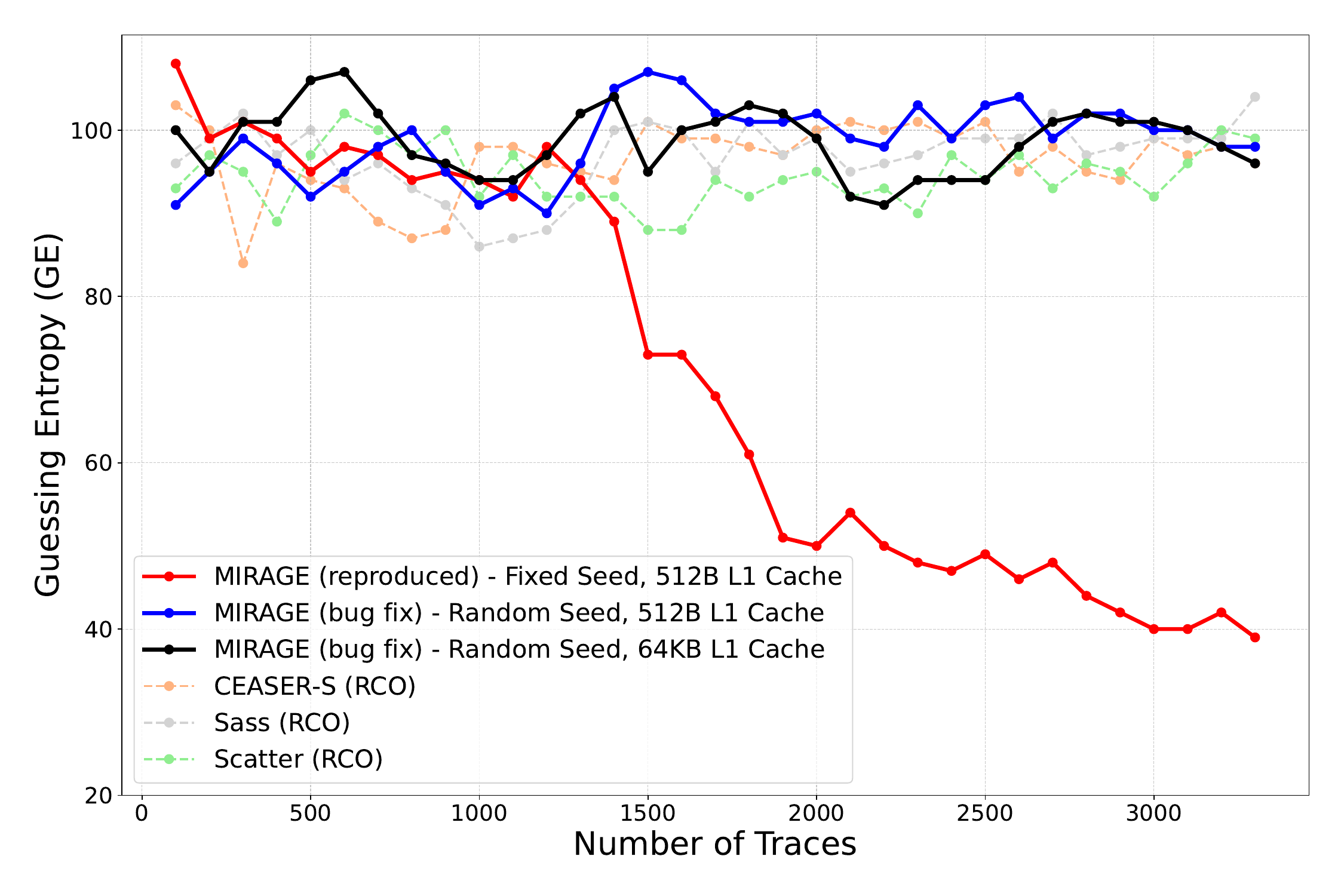}
         \caption{\textbf{Guessing Entropy (GE) for an unknown AES key, after our fixes}, as number of AES encryptions increases.  MIRAGE has a high GE above 90\% even after thousands of AES encryptions, showing no leakage, once the Eviction RNG is initialized with a Random Seed, with both a 512B L1 Cache (like RCO's artifact) and a more realistic 64KB L1 Cache.}
    \label{fig:final_reproduction}
\end{figure}

We evaluate the Guessing Entropy (GE) for a victim's AES key, using the template for a profiled AES key similar to RCO, using the formula: 
$GE = \sum_{i=0}^{15} log_2(R_i)$, where $R_i$ is the rank of the correct guess for key byte $i$. 
\cref{fig:final_reproduction} shows the 
guessing entropy (GE) for Mirage based on their artifact (\textbf{reproduced}), and with our fixes of RCO's modeling issues where we use a random seed to initialize the RNG for global evictions (\textbf{bug fix}). 
After our bug fixes, we see that \textbf{Mirage has high GE above 90\%}, even after thousands of AES encryptions, demonstrating that it is \textbf{resilient to brute-force key guessing attacks on AES} when modeled correctly.

\section{Concluding Remarks}
Our analysis shows that, when modeled faithfully, MIRAGE remains resilient to occupancy-based side-channel attacks aiming to recover AES keys. The results reported in the RCO paper~\cite{chakraborty2025occ} arise from unrealistic modeling, most notably the use of a deterministic sequence of global evictions, across multiple AES encryptions, which do not reflect MIRAGE’s design. We encourage the authors of the RCO paper~\cite{chakraborty2025occ} and the SoK paper on randomized caches~\cite{bhatlasok} to revisit their conclusions in light of these findings.
Our code is open-sourced at \url{https://github.com/sith-lab/yet-another-mirage-of-breaking-mirage}.
\section{Acknowledgment}
We thank Moinuddin Qureshi and Tom Ristenpart for their helpful feedback on an earlier draft of this work.

\bibliographystyle{IEEEtran}
\bibliography{ref}

\appendix
\begin{appendices}

\section{Bug-1: Guessing Entropy Analysis Bug in RCO}\label{app:bug1}
\lstset{aboveskip=0pt, belowskip=0pt}

\begin{figure}[h]
\noindent
\begin{minipage}[t]{0.43\columnwidth}
\begin{tcolorbox}[enhanced,
  sharp corners,
  colback=white,
  colframe=black!40,
  title=Bug\strut,
  fonttitle=\bfseries,
  colbacktitle=black,
  title style={minimum height=3ex},
  boxrule=0.3pt,
  equal height group=bugfix]   
\begin{lstlisting}[language=Python, basicstyle=\ttfamily\tiny,escapechar=!]
def develop_tuples(
    round_timing_tuples, 
    filename, key_position):
    datalines = open(filename).
                    readlines()

    !\colorbox{diffrem}{index = 0}! 
    for data in datalines:
        if(!\colorbox{diffrem}{index}! > max_traces):
            continue
        !\colorbox{diffrem}{index = index + 1}!
        .. % process trace




for filename in all_files:
      develop_tuples(..)
   
\end{lstlisting}
\end{tcolorbox}
\end{minipage}\hfill
\begin{minipage}[t]{0.58\columnwidth}
\begin{tcolorbox}[enhanced,
  sharp corners,
  colback=white,
  colframe=black!40,
  title=Fix\strut,
  fonttitle=\bfseries,
  colbacktitle=black,
  title style={minimum height=3ex},
  boxrule=0.3pt,
  equal height group=bugfix]  
  \begin{lstlisting}[language=Python, basicstyle=\ttfamily\tiny, escapechar=!]
def develop_tuples(
    round_timing_tuples, 
    filename, key_position, !\colorbox{diffadd}{traces\_used}!):
    datalines = open(filename).
                    readlines()

    for data in datalines:
        if(!\colorbox{diffadd}{traces\_used}! > max_traces):
            continue
        !\colorbox{diffadd}{traces\_used = traces\_used + 1}!
        .. % process trace
    return !\colorbox{diffadd}{traces\_used}!

!\colorbox{diffadd}{traces\_used = 0}!
for filename in all_files:
      !\colorbox{diffadd}{traces\_used}!
      = develop_tuples(.., !\colorbox{diffadd}{traces\_used}!)

\end{lstlisting}
\end{tcolorbox}
\end{minipage}
\caption{Bug in RCO's Guessing Entropy Code}
\label{lst:bugfix_rco1}
\end{figure}

\cref{lst:bugfix_rco1} illustrates the bug in RCO's Guessing Entropy Analysis code. In the \textbf{original implementation (left)}, RCO processes multiple trace files (up to 6 AES trace files) to calculate the guessing entropy, creating tuples for computing guessing entropy. The bug arises because the check against \texttt{max\_traces} is applied locally per trace file, rather than globally across all trace files. As a result, the code uses  ends up using $\texttt{6} \times \texttt{max\_traces}$, instead of the intended \texttt{max\_traces}, causing the  number of AES traces needed for a given GE to be under-reported by a factor of six. 

In our \textbf{bug fix (right)}, we introduce a \texttt{traces\_used} variable that tracks the total number of traces used across all files. This ensures that the \texttt{max\_traces} limit is enforced globally correcting this counting error.

Our reproduction stored all the traces in a single file and thus was not affected by this bug. The discrepancy between the reproduced results and RCO’s reported results helped us identify this issue. The code for this bug fix is available \href{https://github.com/sith-lab/yet-another-mirage-of-breaking-mirage/commit/31c6f80}{here}.

\section{Bug-2: Global Eviction RNG Seed Bug in RCO}\label{app:bug2}
\begin{figure}[h]
\noindent
\begin{minipage}[t]{0.48\columnwidth}
\begin{tcolorbox}[enhanced,
  sharp corners,
  colback=white,
  colframe=black!40,
  title=Bug\strut,
  fonttitle=\bfseries,
  colbacktitle=black,
  title style={minimum height=3ex},
  boxrule=0.3pt,
  equal height group=mirage] 
\begin{lstlisting}[language=C++, basicstyle=\ttfamily\tiny,escapechar=!]
VwayTags(const Params *p)
     :BaseTags(p),..
     replacementPolicy(
        p->replacement_policy)
     !\colorbox{diffrem}{mt\_rand(42) // constant seed}!
{ ..
}
\end{lstlisting}
\end{tcolorbox}
\end{minipage}\hfill
\begin{minipage}[t]{0.52\columnwidth}
\begin{tcolorbox}[enhanced,
  sharp corners,
  colback=white,
  colframe=black!40,
  title=Fix\strut,
  fonttitle=\bfseries,
  colbacktitle=black,
  title style={minimum height=3ex},
  boxrule=0.3pt,
  equal height group=mirage] 
\begin{lstlisting}[language=C++, basicstyle=\ttfamily\tiny,escapechar=!]
VwayTags(const Params *p)
     :BaseTags(p),..
     replacementPolicy(
        p->replacement_policy),
{     
  !\colorbox{diffadd}{std::random\_device rd; }!
  !\colorbox{diffadd}{mt\_rand.seed(rd()); // random seed}!
}
\end{lstlisting}
\end{tcolorbox}
\end{minipage}
\caption{Bug in RCO's modeling of MIRAGE's Evictions}
\label{lst:bugfix_rco2}
\end{figure}

\cref{lst:bugfix_rco2} illustrates the bug in RCO's modeling of MIRAGE's evictions. In the \textbf{original code (left)}, the RNG used for global tag evictions is initialized with a \textbf{fixed seed} (42). Due to RCO's attack methodology of restarting the simulation for each AES encryption, this causes every AES simulation to artificially produce the same eviction sequence. 
The \textbf{bug fix (right)} replaces the fixed seed with a random seed  unpredictable to an attacker generated via \texttt{std::random\_device}, producing a \textbf{different global eviction sequence} for each AES simulation. This models a realistic behavior of an attack on MIRAGE.
Our patch for the bug fix is available at this \href{https://github.com/sith-lab/yet-another-mirage-of-breaking-mirage/blob/main/src_randseed_patch/vway_tags.cc.patch}{link}.

\end{appendices}

\end{document}